\documentstyle[12pt]{article}
\begin{document}
\baselineskip .3in
\pagestyle{plain}
\newpage
\begin{center}
{\Large{\bf An extended pair tunneling model: Studies on bilayer 
            splitting and some superconducting state properties}}  
\end{center}
\vskip 1.0cm
\begin{center}
{\bf A. N. Das$^\star$ and Biplab Chattopadhyay$^\dagger$} 
\end{center}
\vskip 0.50cm
\begin{center}

 {\em Saha Institute of Nuclear Physics, \\
1/AF Bidhannagar, Calcutta 700 064, India}\\

\end{center}

\vskip 3.0cm

\indent~~~~~~~~~~ Photoemission intensity. 
\vfill
\hrule   
\noindent $^\star$e-mail: atin@cmp.saha.ernet.in; 
          $^\dagger$e-mail: biplab@cmp.saha.ernet.in 
\newpage
\begin{center}
{\bf Abstract}
\end{center}
\vskip 0.5cm
We consider an extended version of the pair tunneling model
including interlayer single particle hopping (ISPH) as a
complementary process to pair tunneling. The normal state gap,
as found in cuprates, is taken to suppress the effective ISPH
in conformity with the experimental observations, and this in turn
enhances the pair tunneling process. The effective ISPH involves
a probability factor $P$ for which we consider two choices  
and provide phenomenological arguments in favour of them. We address 
the issue of bilayer splitting by calculating the spectral density 
function and corresponding photoemission intensity curves and show 
that our calculations conform with the absence of bilayer splitting 
observed in ARPES experiments on Bi2212. We have also studied the 
temperature variation of the superconducting gap and ratio of the 
superconducting gap to $T_c$. Our results, obtained for both the 
choices of $P$, are reasonably in good agreement with those from 
experiments on cuprate superconductors. A linear $T$-dependent 
choice of $P$, however,  yields a precise match to the experimantal 
data of the temperature varying superconducting gap.

\newpage
\begin{center} 
{\large{\bf I. Introduction}}  
\end{center} 

In order to explain various unusual properties of high-$T_c$
superconductors (HTS) a number of phenomenological models
have been proposed. One such model is the interlayer pair
tunneling model (ILPT), originally proposed by P. W. Anderson
\cite{and1}. The model is based on the postulate that the
coherent single particle tunneling between two $CuO_2$ layers
of HTS is blocked owing to strong electronic correlation effect,
whereas tunneling of Cooper pairs between the layers is allowed.
This delocalization process of the pairs gives rise to a substantial
enhancement of pairing in each layer and raises the transition
temperature \cite{and2}. An important outcome of the ILPT model
is that it yields considerably higher $T_c$ for a bilayer system
compared to that for a single layer system, even with a weak or
moderate in-plane attractive interaction. Thus, the model can
naturally explain higher values of $T_c$ in bi-layered or
triple-layered systems compared to those of monolayer (in
reality well separated layers) systems and obviates the need
for a strong in-plane attractive interaction to achieve
high $T_c$.

In the ILPT model of Anderson \cite{and1,and2} the coherent
single particle tunneling
between layers is assumed to be completely blocked, even though
the bare (interlayer) hopping rate, as obtained from the electronic
structure calculation is substantial \cite{band}.
The idea of absence of interlayer single particle tunneling
originates from the anomalous c-axis transport properties of
cuprates. For underdoped systems rapid increase in the c-axis
resistivity with decreasing temperature and absence of a Drude peak
in the optical conductivity along the c-axis \cite{tamabat}
indicate absence of coherent single particle hopping between the
layers \cite{and2}. Anderson argued that these anomalies
are due to the strong electronic correlations present in cuprates
and are characteristics of a non-fermi liquid behaviour.
However, for sufficiently overdoped cuprates the $c-$axis
resistivity shows metallic like temperature dependence as that in
the $ab-$plane and the Drude peak appars in the ($c-$axis) optical
conductivity spectrum \cite {gray,uchida}.
These facts suggest presence of interlayer single particle
hopping (ISPH) in case of overdoped systems.
In fact, it is generally beleived that
the {\it confinement} of the electrons in CuO$_2$ planes 
is a characteristic feature of underdoped cuprates where correlation 
has its strongest effect. For overdoped systems, on the other  
hand, the effect of correlation becomes weaker, hence the charge 
carriers are expected to behave more like a Fermi liquid. 

Recently, it has been established that a normal state pseudo 
gap ($E_g$) exists in high-$T_c$ systems \cite{loeser,ding}. 
This gap is momentum dependent and is given by 
\begin{equation}
E_g(k) = \frac{1}{2} E_g |Cosk_x a-Cosk_y a|
\end{equation}
where $a$ is the in-plane lattice constant and 
the half factor is introduced so that the maximum 
value of the gap for any particular doping becomes $E_g$.
For underdoped systems $E_g$ is large, it decreases gradually with 
doping and becomes small or negligible for overdoped systems
\cite{loeser,ding, gins}. An interesting correlation has been observed
between the normal state gap and the $c$-axis transport in cuprates.  
For underdoped system, where $E_g$ is large, the interlayer single
particle hopping is negligible, while, for overdoped system, where
$E_g$ is small, it is allowed. Hence, a relationship, that the normal
state gap is connected with the suppression of the interlayer single
particle hopping, could be inferred.

Motivated by the above observations and inference, recently we
proposed an extended pair tunneling model \cite{dasil, bipdas}  
involving both ISPH and interlayer pair tunneling
\cite{dasil, bipdas}, which is an extension
of Anderson's ILPT model \cite{and2}.  
We assume that the particles can tunnel between two layers via
two channels: (i) single particle hopping and (ii) pair
tunneling, and these two processes are complementary. 
We introduced a probability factor ($P$) for the ISPH which 
decreases very fast with the increase of $E_g$ \cite{dasil}.
As a result, for underdoped systems where $E_g$ is 
large, the ISPH is practically blocked and the only relevant 
interlayer coupling is through the pair tunneling. Our proposed model 
thus becomes equivalent to Anderson's ILPT model \cite{and2} for highly
underdoped systems.
For the overdoped case, where $E_g$ is small, the ISPH within our
extended model becomes appreciable
with the reduction of pair tunneling process. Thus, the extended
model is consistent with the implications of the transport data
which predict presence of ISPH for sufficiently overdoped systems.
Certain important results of the extended ILPT
model are:
(i) it can naturally give high values of the ratio
of the superconducting gap to $T_c$,
(ii) with increasing interlayer coupling $T_c$ may increase
or decrease depending on the relative dominance of pair tunneling
or ISPH, 
(iii) with the doping and other parameters remaining unchanged, 
$T_c$ becomes higher for materials with larger $E_g$,
(iv) temperature varying superconducting gap data is in closer agreement
with that within the extended ILPT model than the original ILPT model. 
These results are similar to those characterizing the high-$Tc$ cuprate
superconductors. Some of these have already been presented
elsewhere \cite{dasil, bipdas}.

The presence of ISPH in our model may raise 
certain relevant questions regarding the issue of bilayer splitting
in cuprate superconductors.
The ISPH, in general, gives rise to splitting  
of bands which should be reflected in the electronic 
density of states (DOS) as well as in the spectral function, 
as the presence of a two peak struture. 
However, in recent angle resolved photoemission spectroscopy 
experiments (ARPES) \cite{ding2},  
no evidence for bilayer splitting is observed in Bi2212 system. 
Therefore, it becomes important to investigate whether such 
observation is contradicted within our proposed extended model. 
In this communication we particularly focus on this issue. 
By a detailed study of the ARPES intensity curves, we show that 
the absence of bilayer splitting, as observed in experiments, 
can be understood within our proposed extended model. 
Our calculations are carried out including realistic band 
structure of Bi2212 \cite{norman} and the results are compared 
with the experimental data.

In the proposed extended ILPT model \cite{dasil,bipdas} we
consider the effective ISPH along with a probability
factor $P$, which is suppressed by $E_g$. Since the microscopic
origin of $E_g$ is still unknown and the proper functional dependence
of $P$ on $E_g$ is difficult to predict, we make two choices for $P$.
In the first one \cite{dasil} we consider
\begin{equation}
P= {\rm e}^{-E_g/T}
\end{equation}
which is henceforth referred to as the {\it exponential form}.
In another form \cite{bipdas}
\begin{equation}
P = \frac{T}{E_g+T {\rm e}^{-E_g/T}}
\end{equation}
which will be called as {\it $T$-linear form}.
The exponential form could signify that the charge carriers
has to overcome a gap for being available for the ISPH,
whereas the $T$-linear
form signifies no gap, but a simple power law dependence of $P$ on 
$T$ and $E_g$. The latter form could find a justification within
the spin-charge separation picture. Towards the end, we present
arguments regarding these choices of $P$.

In this communication we also make a comparative study of the
properties related to the superconducting gap with the above
mentioned two different choices of the probability factor $P$.
We find that, in general, the results are promising for both 
the choice, but the $T$-linear choice yields a better match to 
the experimental gap-variation data \cite{expgap} than the 
exponential choice. 

\vskip 1.5cm

\begin{center} 
{\large{\bf II. Formalism}}  
\end{center} 
\vskip 0.5cm

The model hamiltonian for the coupled bilayer system is given by 
\cite {dasil, bipdas}
\begin{eqnarray}
H = H_N + H_S
\end{eqnarray} 
where  \begin{eqnarray}
H_N= \sum_{i,k,\sigma} (\epsilon_k - \mu) c_{k \sigma}^{(i)+} 
c_{k \sigma}^{(i)}  + \sum_{i\neq j,k,\sigma} \{ t_{\perp}^{\rm eff}(k) 
c_{k \sigma}^{(i)+} c_{k \sigma}^{(j)} + h.c \} 
\end{eqnarray} 
\noindent 
and \begin{eqnarray} 
H_S &=&  -\sum_{i,k,k'} \{ V_{k,k'} c_{k \uparrow}^{(i)+} 
       c_{-k \downarrow}^{(i)+} c_{-k' \downarrow}^{(i)} 
       c_{k' \uparrow}^{(i)} + h.c \}  \nonumber   \\
  &-& ~\sum_{i\neq j, k} \{ T_p^{\rm eff}(k) c_{k \uparrow}^{(i)+} 
c_{-k \downarrow}^{(i)+} 
c_{-k \downarrow}^{(j)} c_{k \uparrow}^{(j)} + h.c \} 
\end{eqnarray}

The operator $c_{k \uparrow}^{(i)+}~(c_{k \uparrow}^{(i)})$ is the 
fermion creation (annihilation) operator with momentum $k$, 
spin $\uparrow$ and in the layer $i (=1,2)$. 
$H_N$ describes the band energy of the bilayer system, where 
$\epsilon_k$ is the band dispersion in a layer,  
$t_{\perp}^{\rm eff}$ is the effective interlayer single particle 
hopping matrix element and $\mu$ is the chemical potential. $H_S$ 
represents the interaction part of the hamiltonian and it leads 
to superconductivity in the system below the transition temperature 
$T_c$. $V_{k,k'}$ is the in-plane pairing 
interaction which is responsible for the formation of Cooper 
pairs in a layer. $T_p^{\rm eff}$ represents the effective interlayer 
pair tunneling matrix element. 
For the pair tunneling term we have  considered  
only the diagonal pair hopping \cite{and2}, 
where quasi particle momentum is conserved during tunneling.  
Effective ISPH 
is taken to be $t_\perp^{\rm eff}(k) = t_\perp^b(k) P$, where 
$t_\perp^b(k)=t_\perp((\cos k_xa-\cos k_ya)/2)^2$ 
is the $k$-dependent ISPH as predicted by the band structure 
calculations \cite{band} with $t_\perp$ being the bare ISPH 
matrix element, and $P$ is the ISPH probability factor for which 
we consider the two forms as given in Eqs. (2) and (3).
Pair tunneling, quite naturally, should be connected to the 
ISPH probability, since only those particles which are not 
taking part in ISPH, are available for the process.
Tunneling of pairs, is a two particle process, hence, it 
occurs with a probability $(1-P)^2$ \cite {dasil}.  
The effective pair tunneling matrix element is, thus,  
given by $T_p^{\rm eff}(k) = T_p(k) \,(1-P)^2$. The (bare) 
pair tunneling, following Ref. \cite{and2}, is taken as   
$T_p(k) = [(t_\perp^b(k))^2/|t_1|]$, where 
$t_1$ is the nearest neighbour hopping matrix element of the 
$ab$-plane band dispersion. It may be
noted that for large $E_g$ or at very low temperatures the ISPH
probability factor $P$ becomes very small within our model and
all the results predicted by our model become same as that of
Anderson's ILPT model. In fact, in the limit ${E_g/T \to \infty}$
(${P \to 0}$), our model reduces exactly to the original ILPT model
by Anderson. 

Mean field decoupling of four fermion terms in $H_S$ yields 
\begin{eqnarray}
H_S &=& - \sum_{i,k} \left[\Delta_{k}\, c_{k \uparrow}^{(i)\dag} 
      c_{-k \downarrow}^{(i)\dag} + h.c \right] 
\end{eqnarray}
where $\Delta_{k}$ is the superconducting gap and is given by  
\begin{equation}
\Delta_k = \Delta_{i,k} = \sum_{k^\prime} 
  V_{k,k^\prime}\, \langle c_{-k^\prime \downarrow}^{(i)} 
  c_{k^\prime \uparrow}^{(i)}\rangle  
  + T_p^{\rm eff}(k)\, \langle c_{-k \downarrow}^{(j)} 
  c_{k \uparrow}^{(j)}\rangle 
\end{equation}
The layers $i$ and $j$ are equivalent, since by symmetry, the in-plane
pairing average is identical in both the layers. 

It is well known that presence of interlayer single particle hopping 
between two layers produces bonding and antibonding bands.  
The corresponding annihilation operators are defined as
  $  c_{k \uparrow}^{(-)}=\frac{1}{\sqrt{2}} (c_{k \uparrow}^{(1)}
			-  c_{k \uparrow}^{(2)})$
and  $  c_{k \uparrow}^{(+)}=\frac{1}{\sqrt{2}} (c_{k \uparrow}^{(1)}
			+  c_{k \uparrow}^{(2)})$
and    
$H_N$ becomes diagonal in this new representation. The full 
hamiltonian may be written in terms of these new operators as  
\begin{eqnarray}
H &=& \sum_{i} \left[ \sum_{k,~\sigma}  \xi_k^{(i)}
      c_{k \sigma}^{(i)\dag} c_{k \sigma}^{(i)} - 
      \sum_{k}  (\Delta_{k}\, c_{k \uparrow}^{(i)\dag} 
      c_{-k \downarrow}^{(i)\dag} + h.c) \right] 
\end{eqnarray}
where $i$ is now the band index and $i=~-~(+)$ represents the bonding 
(antibonding) band. Here,  
 $  \xi_ k^{(\pm)}  = (\epsilon_k - \mu) \pm t_\perp^{\rm eff}(k)$ 
 are the normal state band energies. From the hamiltonian of Eq.(9)
the self consistent equations for the chemical potential and the 
superconducting gap are obtained as 
\begin{equation}
1-\delta = 1-\frac{1}{N} \sum_k \xi_k^{(-)} \chi(E_k^-)
	 - \frac{1}{N} \sum_k \xi_k^{(+)}  \chi(E_k^+)
\end{equation}
and 
\begin{equation}
\Delta_k = \frac{\displaystyle\sum_{k^\prime} \Delta_{k^\prime}\, 
		V_{k,k^\prime} 
		\left(\chi(E_{k^\prime}^-)+
		\chi(E_{k^\prime}^+)\right)\!\!\bigg/2}
	       {1 - T_p^{\rm eff}(k)                         
		\left(\chi(E_k^-)+\chi(E_k^+)\right)\!\!\bigg/2} 
\end{equation}
where
$E_ k^{\pm} =\sqrt{ {\xi_k^{(\pm)}}^2 + \Delta_k^2}$, 
are the quasiparticle energies in the superconducting state,  
$\chi(E_k^\pm)=\frac{1}{2E_k^\pm}
\tanh\left(\frac{\beta E_k^\pm}{2}\right)$, $\beta = 1/T$ (in 
a scale of k$_B$=1),   
$\delta=1-n$, with n being the number of electrons per site, and 
$N$ is the total number of lattice sites. 
The in-plane attractive pairing interaction is considered here 
only between nearest neighbors for which    
$V_{k,k^\prime}$ is separable as 
$V_{k,k^\prime} = V\,\eta_k\,\eta_{k^\prime}$, which also 
makes the $k$-dependence of $\Delta_k$ to be separable. 
In our calculations, the $d_{x^2-y^2}$ symmetry of the pairing state 
is considered because of growing evidence of the same in high-$T_c$ 
cuprates \cite{dwpap1,dwpap2}. This implies 
$\eta_k = (\cos k_xa -\cos k_ya)/2$.

Finally, the equation for the superconducting gap is obtained as 
\begin{equation}
{1\over {4V}} = \frac{1}{N} \sum_{k} 
	     \frac{\eta_k^2\,\left(\chi(E_k^-) 
	     + \chi(E_k^+)\right)\!\!\bigg/2} 
	     {1 - T_p^{\rm eff}(k)\left(\chi(E_k^-) 
	     + \chi(E_k^+)\right)\!\!\bigg/2} 
\end{equation}
where the parameter $V$ is the nearest neighbour in-plane attractive 
interaction strength.  

The ARPES intensity is proportional to 
the product of the spectral density function and the fermi 
distribution function. 
The spectral function in the superconducting state  
is given by
\begin{equation}
A(k,\omega) = A^-(k,\omega) + A^+(k,\omega)
\end{equation}
which includes contributions from both the (bonding and antibonding)
quasiparticle bands.
Spectral functions for the two bands are given by 
\begin{equation}   
A^{\pm}(k,\omega) = \frac{1}{\pi} \left [\frac { {(u_k^{\pm})}^2 \Gamma} 
{ {(\omega-E_k^{\pm})}^2 + \Gamma^2}
 + \frac { {(v_k^{\pm})}^2 \Gamma} { {(\omega+E_k^{\pm})}^2 
 + \Gamma^2} \right] ,
\end{equation}
where
${(u_k^{\pm})}^2 =1-{(v_k^{\pm})}^2 =~ \frac{1}{2} 
		  \left ( 1+\frac{\xi_k^{\pm}}{E_k^{\pm}} \right)$
and $\Gamma$ is a phenomenological linewidth parameter accounting
for the broadening of the quasiparticle states due to finite lifetime.

Considering the experimental energy resolution factor the intensity of 
the ARPES is given by \cite{ding3}
\begin{equation}
I(k,\omega)= I_0  \int_{-\infty}^{\infty} R(\omega-\omega^{\prime}) 
f(\omega^{\prime}) A(k^{\prime},\omega^{\prime}) d\omega^{\prime}
\end{equation}
where 
$R(\omega)$ is the Gaussian energy resolution function and 
$f(\omega)~=1/(e^{\beta \omega} +1)$ is the fermi distribution 
function.

\vskip 1.5cm

\begin{center} 
{\large{\bf III. Results and Discussion}} 
\end{center} 
\vskip 0.5cm
A relevant point of contention, that may be raised regarding the 
validity of our model, as proposed in Ref. \cite{dasil, bipdas},
is that in experiments no evidence of
bilayer splitting is observed in Bi2212. This suggests that the   
interlayer single particle hopping would be absent, while a finite 
ISPH is present in our model. So, at first we present the results   
about the ARPES intensity within our model.
By numerical solution of the self-consistent equations (10) and
(12) we evaluate the superconducting gap as a function of
temperature and hence determine the spectral density 
function as well as the ARPES intensity in the superconducting 
phase. For  
the in-plane band dispersion we consider a six parameter tight 
binding band structure, $[t_0,t_1,t_2,t_3,t_4,t_5]=\,[0.131,
-0.149,0.041,-0.013,-0.014,0.013]$ $eV$, where
$t_0$ is the bare Wannier orbital energy, $t_1$ nearest neighbor,
$t_2$ next nearest neighbor etc. hopping matrix elements, 
as described \cite{norman} and used
in previous publications \cite{biplab}. 
We take $t_{\perp}=40\,meV$ and the in-plane attractive interaction 
$V=70\,meV$, which are reasonable. To begin with, we consider
the ARPES intensity and single particle DOS for the exponential
choice of $P$. 

The ARPES intensity $I(k,\omega)$ has been calculated using 
Eqs.(13)-(14) at the Brillouin zone point $k=(\pi,0)$
for optimal doping and for different values of the energy
resolution characterised by the full width at half
maximum (FWHM) values of the Gaussian energy resolution
function. The value of $E_g$ at optimal doping is taken from 
a best fit analysis of the theoretical curve for the temperature 
variation of the superconducting gap to the experimental data, 
which is presented later.
The plots of $I(k,\omega)$ versus $\omega$ at different temperatures 
are shown in Fig.1, where we take a very small value for the
linewidth parameter $\Gamma \sim 1\,meV$. A perfect  
energy resolution is assumed in Fig.1a. The figure shows that even 
within the limit of perfect resolution (FWHM=0) no 
bilayer splitting could be observed at T=13K, while at T=40K a 
splitting is observable in $I(k,\omega)$, but the difference 
in energy between the two peaks is only $\sim 8 \,meV$ . 
Fig. 1b shows the evolution of $I(k,\omega)$ at T=40K from a two 
peak structure to a single peak structure as the FWMH of the 
energy resolution function increases 
from $2\,meV$ to $10 \,meV$. The experimental energy resolution 
for the ARPES, as quoted by different workers \cite{ding3,shen}, 
is $\sim 19 \,meV$, which is larger than that we have
considered. Furthermore, at $T=40\,K$ the linewidth
parameter in cuprates would be much larger than
$1\,meV$ \cite{exprat}, which will further broaden the spectral
function and hence the $I(k,\omega)$. So, it is clear that 
no bilayer splitting could be observed in 
the ARPES experiments within our model for the 
parameter space, relevant for high-$T_c$ cuprates, and for the 
exponential choice of $P$. 

Presence of the ISPH in a bilayer system would also produce a 
splitting in the normal state electronic DOS.  
So a study of the normal state DOS could provide a  
way to examine the significance of the bilayer splitting. 
In Fig.2  we plot the electronic DOS at different temperatures 
and for different values of the linewidth parameters $\Gamma$. 
It may be noted that the effective ISPH in our model is temperature 
dependent and it makes the DOS to be dependent on temperature. 
Fig.2a shows the plot of the DOS at $T$=10K. It is found that  
even with $\Gamma$=0, no evidence of bilayer splitting in the 
DOS could be observed within our model for the exponential choice
of $P$ at very low temperatures. For $T$= 50K, bilayer splitting is
quite evident for $\Gamma$=0. However, as the value of $\Gamma$ 
is increased to $14 \,meV$ only one broad peak is observed in the 
DOS. 
For high-$T_c$ cuprates the value of $\Gamma$ is quite large  
at high temperatures \cite{exprat}. Thus, our study shows that
bilayer splitting
may not be observable experimentally even if the 
normal state DOS is probed directly. 

In Fig.3 we present the plot of the intensity of the ARPES  
$(k=\pi,0 )$ for the $T$-linear form of $P$ and 
for a value of the linewidth parameter $\Gamma=1\,meV$. For 
this form of $P$ the suppression of the effective ISPH by $E_g$ 
at low temperatures is much less rapid than the exponential form 
of $P$. Consequently, at $T=13K$ the $I(k,\omega)$ shows a two peak 
structure when the energy resolution is perfect (FWHM=0). 
However, even for FWHM as small as 3 $meV$ the $I(k,\omega)$ takes 
a single peak form at $T=13K$. Fig. 3b shows how the two peak form  
of the $I(k,\omega)$ at $T=40\,K$ evolves into a single peak with
decreasing energy resolution or increasing value of FWHM of the
Gaussian energy resolution function.
A single peak form for $I(k, \omega)$ is obtained at $T=40K$ for
a value of $\Gamma$ as small as $1\, meV$ and a value of FWHM of
the energy resolution function as $12 \, meV$, which is less than
the FWHM values of the resolution function in
experiments \cite{ding3,shen}.
It should be noted that consideration
of higher values of $\Gamma$ would produce increased broadening in
the shape of the $I(k,\omega)$, hence a single peak form would be 
obtained even with a better energy-resolution (a lower value
of FWHM) in actual experiments.

In previous works \cite {dasil,bipdas} we studied the superconducting 
phase diagram and some of the superconducting properties
within our model for different $E_g$ values.
The value of $E_g$ changes with doping in cuprates. 
However, for optimal doping $E_g$ has a fixed value for any
particular cuprate material.
In this communication we study different superconducting 
properties only at optimal doping and make a comparison of the 
results, obtained with two choices of $P$, with experimental data.  
In Fig.4 we present the best-fit curves for the temperature 
variation of the superconducting gap for two choices of $P$ 
alongwith the experimental data points. The curve for the 
Anderson's pair tunneling model ($E_g=\infty$) is also shown. 
It is clearly seen that predictions within our model are much 
closer to the experimental data than that with the original 
ILPT model. For the exponential choice of $P$, it is found that 
$E_g \sim 6-7\,meV$ gives a very good fit to the experimental data 
of Bi2212 at optimal doping. 
This value of $E_g$ is very close to the transition temperature of 
Bi2212 system at optimal doping. It is interesting to note that  
the value of $E_g \sim T_c$ at optimal doping has been suggested 
in cuprates from experimental observations also \cite{gins}. 

However, with the exponential form of $P$ we find that it is 
difficult to obtain good fit to the experimental data points for 
the gap variation at low temperatures. A precise fit to the 
experimental data is possible (Fig.4) for the second choice of $P$, 
where it increases almost linearly with temperature (Eq.(3)). 
However, the value of $E_g$, required for such a fit is higher 
than that for the exponential choice of $P$. It should be noted 
that due to the absence of any microscopic derivation for the 
correct form of $P$, we make possible choices which are 
purely phenomenological. In the exponential choice, $E_g$ 
appears as a gap to hinder the charge transport along the 
$c-$axis. But, in the $T$-linear choice there is no direct gap 
and the conditions that $P$ in Eq.(3) has to satisfy, are: 
(i) $P$ increases linearly with $T$ while decreases with $E_g$ 
for $T \ll E_g$ and (ii) $P=1\,$ for $E_g=\,0$. Clearly, these 
conditions could be satisfied even if a prefactor is added to 
$E_g$. Thus the value of $E_g$, as in Eq.(3), may differ from 
the actual value by a constant factor.
					       
In high-$T_c$ systems the ratio of the superconducting gap to $T_c$
is very high compared to the conventional superconductors. The 
value of the gap ratio $(2 \Delta_0 / T_c)$ in cuprates is reported 
to be within the range 6-9  \cite{exprat}. 
In Fig.5 we plot the gap ratio   
as a function of interlayer single particle hopping matrix element 
($t_{\perp}$) for two different values of $E_g$ and for both the
forms of $P$, for a doping level which is optimal for
$t_{\perp} =40\, meV$.
The ratio increases with increasing value of $t_{\perp}$. 
It is seen that both choices of $P$ yield quite high values 
of the gap ratio, but for the $T$-linear choice of $P$
the ratio is larger. For $t_{\perp} =40\, meV$, which is realistic
for layered cuprates, the value of the gap ratio lies within the
range 5.0-7.5 (depending on $E_g$ and the choice of $P$). 
Thus, our model could yield high values for the gap ratio in 
agreement with the experimental findings in cuprates. 

It is observed that the gap ratio is higher for a
smaller value of $E_g$ (Fig.5). Within our model a lower value of
$E_g$ yields a lower value for $T_c^m$ (transition temperature at
optimal doping) \cite{dasil} and this is consistent with the
experimental findings \cite{gins}.
The results of Fig. 5, thus, predict a higher value of the
gap ratio for a cuprate system with lower value of $T_c^m$. 

As found in the results presented above that the qualitative 
characteristics of them do not depend much on the specific 
form of the probability factor $P$ used. While making a choice, 
one only needs to ensure that the probability factor $P$, and as 
a result the effective ISPH, follows the observed doping dependence 
of $c$-axis transport, that is the effective $c\!-\!$axis hopping remains 
heavily suppressed in the underdoped side and at low temperatures. 
In experiments, $E_g$ is found to grow in magnitude \cite{loeser,ding} 
towards underdoing. Thus $E_g$, on phenomenological grounds, could 
be taken to suppress $c$-axis hopping, and hence our choices of $P$ 
could follow. 
 
It should be mentioned that, an exponential suppression of effective 
c-axis hopping towards underdoping was suggested \cite{gray} from 
the experimental results in Y-123 system where a metallic like 
behavior of c-axis resistivity at high temperatures and a 
semiconducting behavior at low temperatures have been observed. 
But, our calculations show that the $T$-linear choice of $P$ gives
a better match to the experimental gap-variation data. One could 
probably search for a realization of this $T$-linear choice within 
within the RVB picture \cite{andzou}. According to the RVB model, 
holons and spinons are the quasiparticles in a layer of strongly 
correlated system and a holon has to combine with a spinon to 
form a hole which can then hop from one layer to another. 
Consequently, $c-$axis hopping is proportional to the spinon 
density, which increases linearly with $T$ within the RVB theory 
and hence, the effective $c-$axis single particle hopping is 
expected to be proportional to $T$.

\vskip 1.5cm
\begin{center} 
{\large{\bf IV. Conclusions}} 
\end{center} 
\vskip 0.5cm
Investigation of the spectral density function and corresponding 
intensity curves for the ARPES within our proposed extendeded  
pair tunneling model shows that the model has no contradiction 
with the absence of bilayer splitting observed in Bi2212 by ARPES 
experiment \cite{ding2}. Studies with two different forms for the 
probability factor $P$, associated with the effective interlayer single 
particle hopping, show that both the forms could yield high values 
for the ratio of the superconducting gap to $T_c$, as observed in 
cuprates. Temperature varying superconducting gap from our 
calculations with both the forms agrees well with the experimental 
data and the agreement is much better compared to that within the 
original ILPT model. A precise fit to the experimental gap variation 
data is achieved with the $T$-linear choice of $P$.
Another prediction of our model is that a system which has
a lower value of $E_g$ should have a lower value of $T_c^m$
and a larger value of the gap ratio. 

It may be mentioned that Anderson's ILPT model is applicable in
the underdoped region where non-fermi liquid behavior
dominates and the ISPH is negligible, whereas our extended model has 
the flexibility of being applicable in a wide region from
the underdoped to a highly overdoped regime where the ISPH is
appreciable.
 
In the model, presented here, we have
considered the effect of $E_g$ only on the interlayer hopping. 
However, it seems natural to think that, $E_g$ should also affect 
the in-plane charge dynamics. In this connection, it has been 
suggested that $E_g$ reduces the in-plane electronic density of 
states \cite{gins}. It would be interesting to study the 
extended model with the pseudogap $E_g$ affecting both the 
in-plane and out of plane charge dynamics. This would constitute 
future communications.

\newpage

\noindent{\Large{\bf Figure captions:}}  

\noindent
Fig.1. Plot of the ARPES intensity $I(k,\omega)$ in arbitrary units 
versus $\omega$ ($eV$) for the exponential choice of $P$. 
Different values of FWHM of the Gaussian energy resolution function
and temperatures are as given in the figure.
The linewidth parameter is taken to be $\Gamma = 1\,meV$.

\vskip 0.5 cm

\noindent
Fig.2. Plot of the electronic density of states $N(\xi)$ 
versus $\xi$ for the exponential choice of $P$. Temperatures
and different values of linewidth parameter $\Gamma$ are mentioned
in the figure.

\vskip 0.5 cm  

\noindent
Fig.3. The ARPES intensity $I(k,\omega)$ versus $\omega$ 
for the $T$-linear choice of $P$ with the linewidth parameter
$\Gamma = 1\,meV$. Values of FWHM and temperatures are listed
in the figure. 

\vskip 0.5 cm
\noindent
Fig.4. Finite temperature gap scaled to its zero temperature value  
  ($\Delta_k^{max}(T)/\Delta_k^{max}(0)$), as a function of 
  reduced temperature ($T/T_c$). Solid line is that in the
  Anderson limit ($E_g = \infty$), solid
  square symbols are experimental data from Ref.\cite{expgap} and 
  dashed lines are from our calculations for two different 
  forms of the probability factor $P$. 

\vskip 0.5 cm 
\noindent
Fig.5. Maximum value of the gap-ratio ($2\Delta_k^{max}(0)/T_c$) 
  versus bare interlayer coupling ($t_\perp$), for different
  $E_g$ values and different choices of $P$, as mentioned in
  the figure.

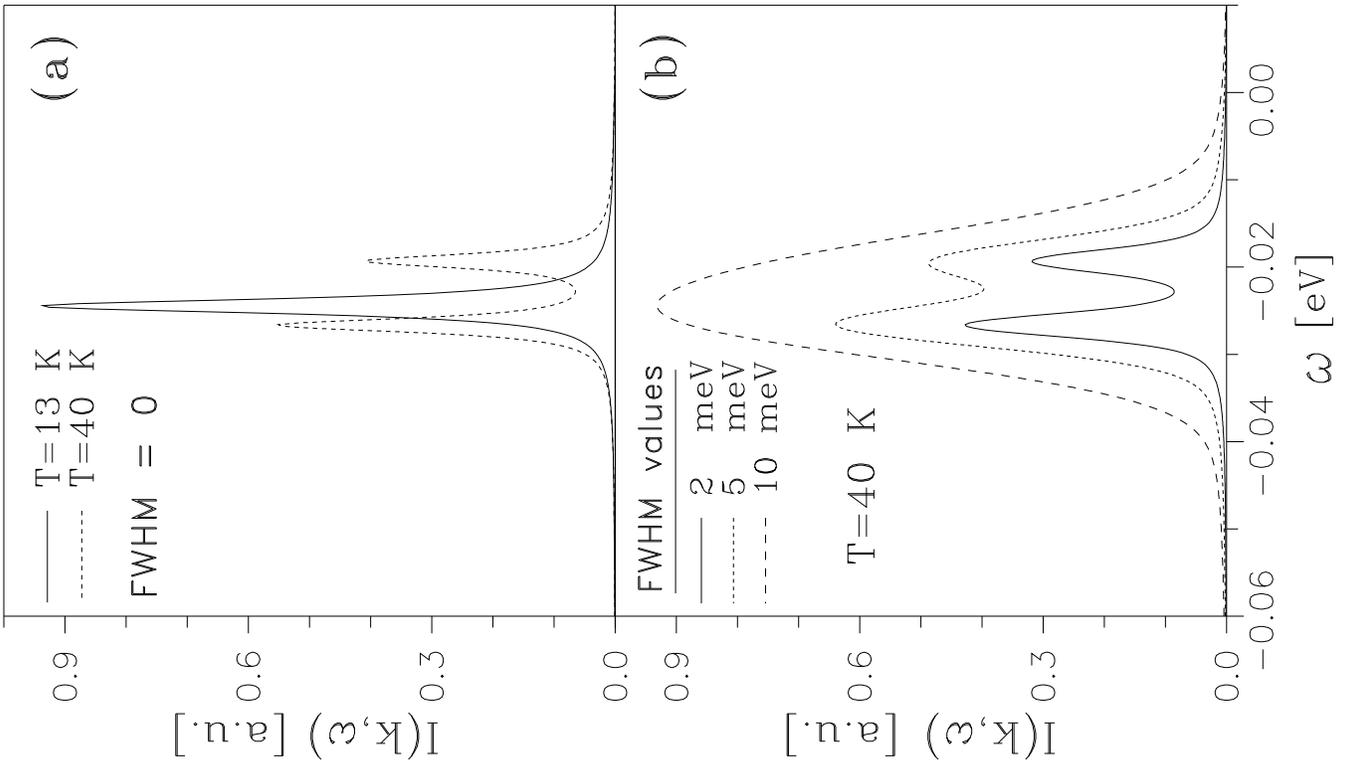
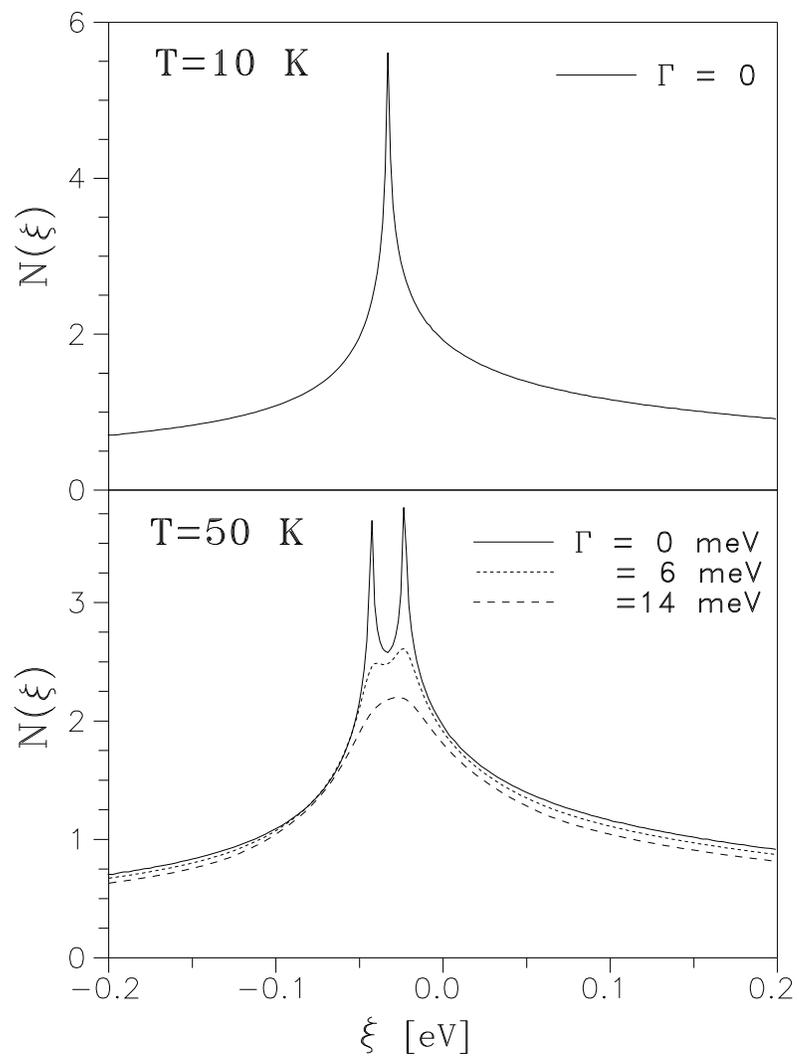
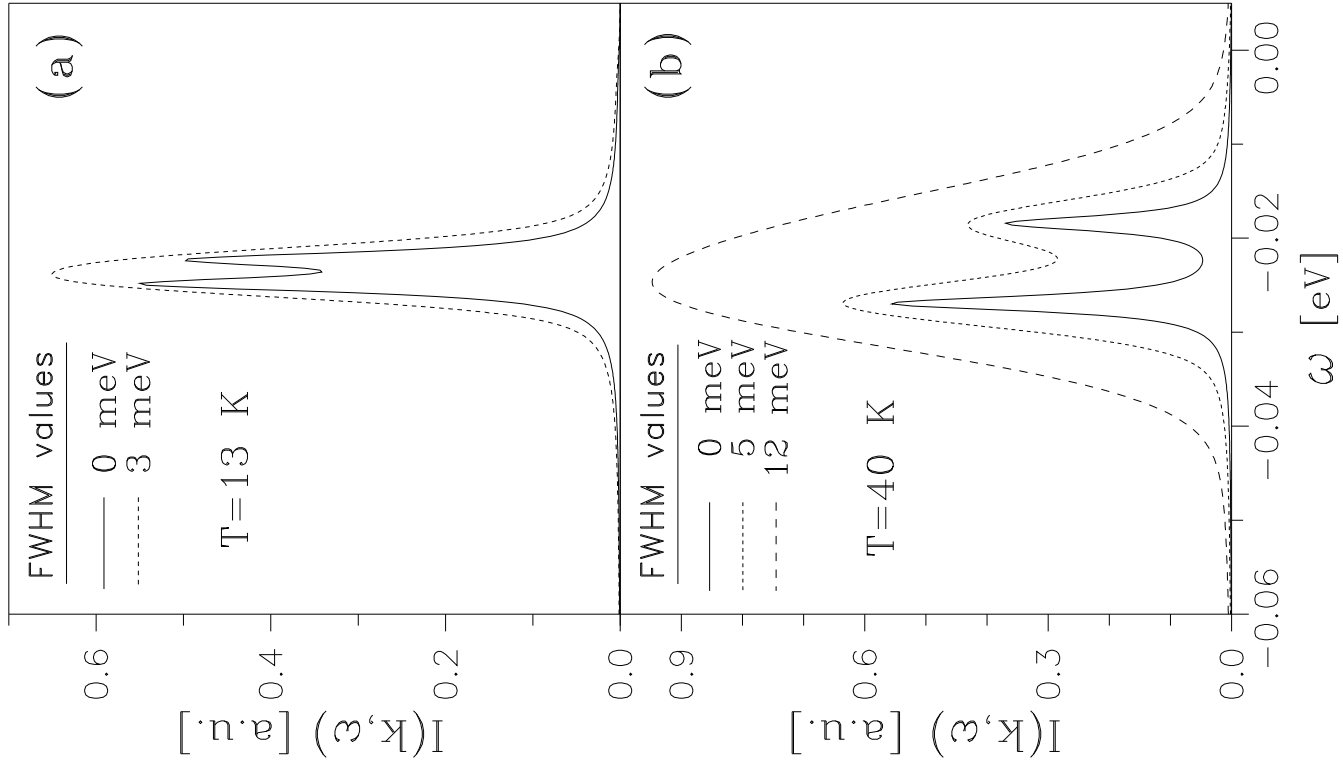
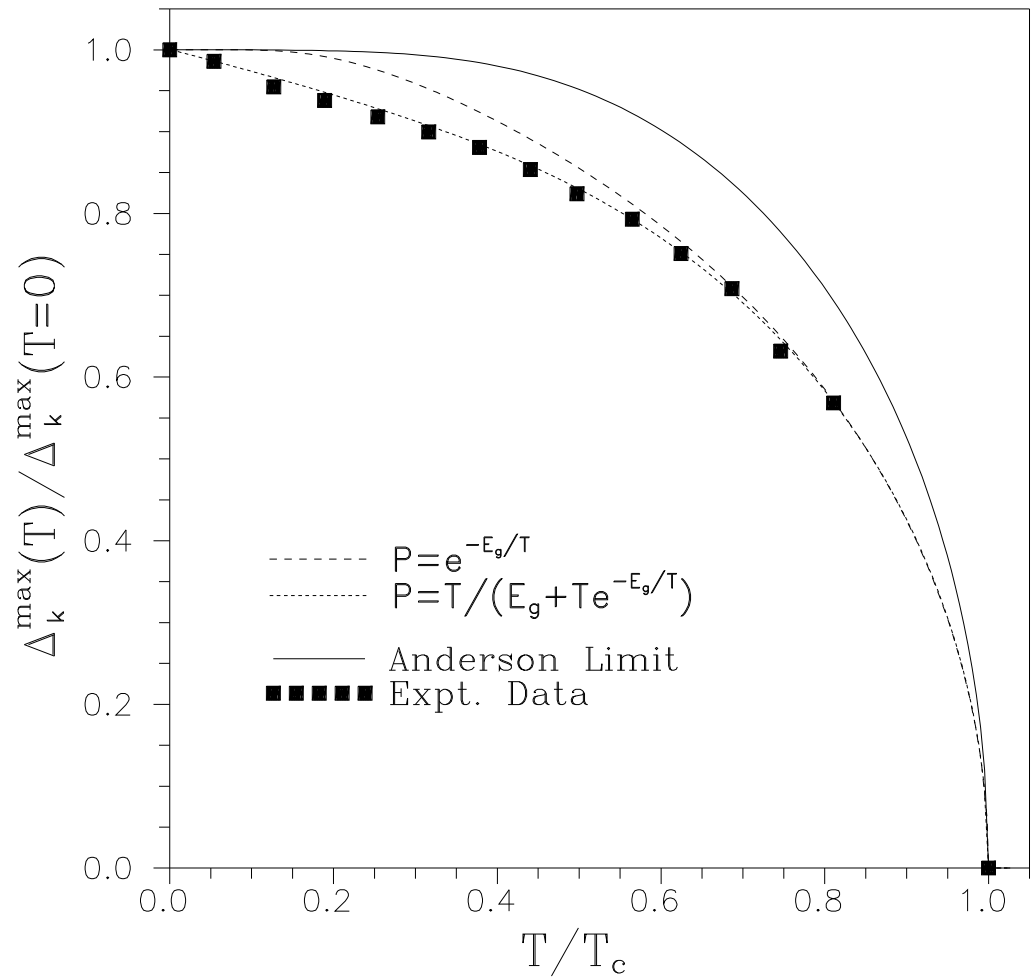
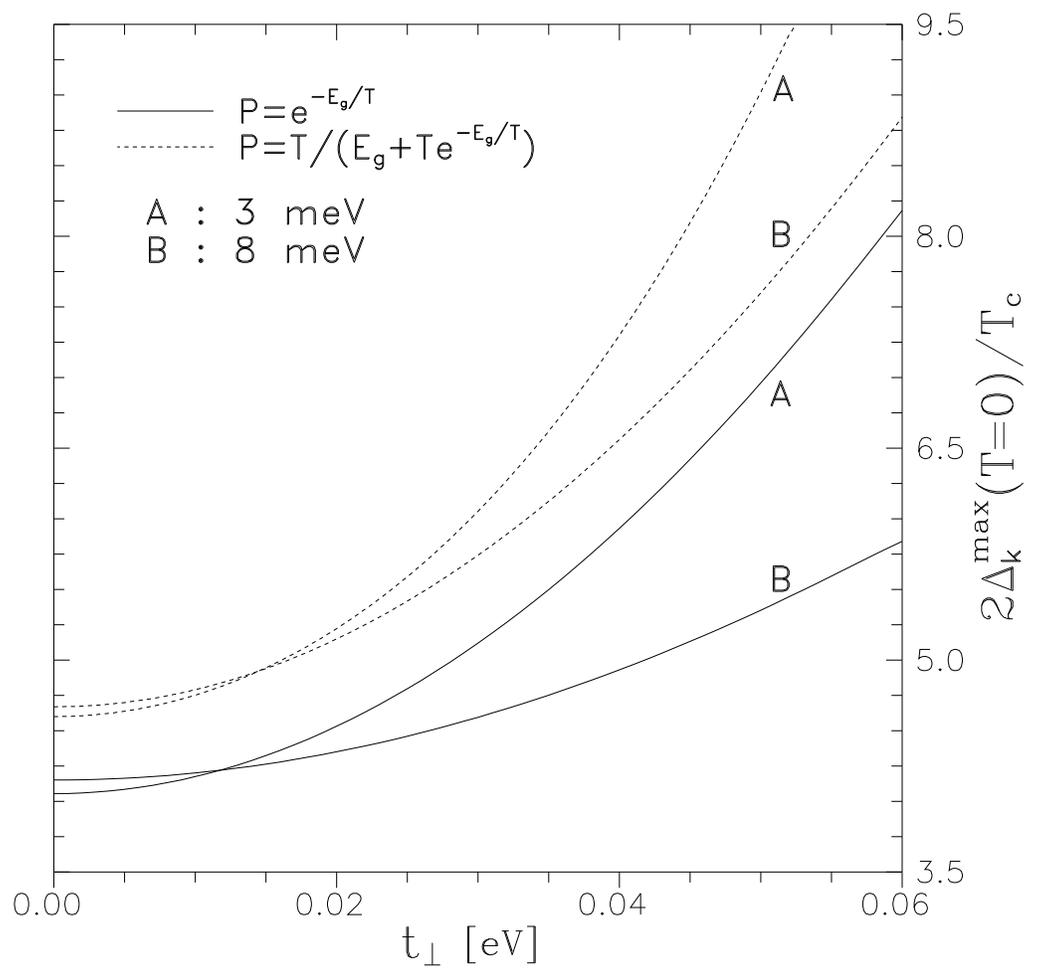
\end{document}